\documentclass[lettersize,journal]{IEEEtran}
\usepackage{amsmath,amsfonts}
\usepackage{algorithmic}

\usepackage{array}
\newcolumntype{M}[1]{>{\centering\arraybackslash}m{#1}}

\usepackage[caption=false,font=normalsize,labelfont=sf,textfont=sf]{subfig}
\usepackage{textcomp}
\usepackage{stfloats}
\usepackage{url}
\usepackage{verbatim}
\usepackage{graphicx}
\def\BibTeX{{\rm B\kern-.05em{\sc i\kern-.025em b}\kern-.08em
    T\kern-.1667em\lower.7ex\hbox{E}\kern-.125emX}}
\usepackage{balance}
\usepackage{cite}

\usepackage{tikz}
\newcommand{\mymkr}[1]{
    \begin{tikzpicture}
    \draw (0,0) node[rectangle,
          rounded corners,
          very thin,
          inner sep=2pt,
          draw=black!10!red,
          ](A){#1};
    \end{tikzpicture}
}

\newcommand{\mymkb}[1]{
    \begin{tikzpicture}
    \draw (0,0) node[rectangle,
          rounded corners,
          very thin,
          inner sep=2pt,
          draw=white!10!blue,
          ](A){#1};
    \end{tikzpicture} 
}
\newcommand{\mymkw}[1]{
    \begin{tikzpicture}
    \draw (0,0) node[rectangle,
          rounded corners,
          very thin,
          inner sep=2pt,
          draw=white!30!white,
          ](A){#1};
    \end{tikzpicture} 
}

\usepackage{scalerel}
\usetikzlibrary{svg.path}
\definecolor{orcidlogocol}{HTML}{A6CE39}
\tikzset{
  orcidlogo/.pic={
    \fill[orcidlogocol] svg{M256,128c0,70.7-57.3,128-128,128C57.3,256,0,198.7,0,128C0,57.3,57.3,0,128,0C198.7,0,256,57.3,256,128z};
    \fill[white] svg{M86.3,186.2H70.9V79.1h15.4v48.4V186.2z}
                 svg{M108.9,79.1h41.6c39.6,0,57,28.3,57,53.6c0,27.5-21.5,53.6-56.8,53.6h-41.8V79.1z M124.3,172.4h24.5c34.9,0,42.9-26.5,42.9-39.7c0-21.5-13.7-39.7-43.7-39.7h-23.7V172.4z}
                 svg{M88.7,56.8c0,5.5-4.5,10.1-10.1,10.1c-5.6,0-10.1-4.6-10.1-10.1c0-5.6,4.5-10.1,10.1-10.1C84.2,46.7,88.7,51.3,88.7,56.8z};
  }
}
\newcommand\orcidicon[1]{\href{https://orcid.org/#1}{\textsuperscript{\scalerel*{
\begin{tikzpicture}[yscale=-1,transform shape]
\pic{orcidlogo};
\end{tikzpicture}
}{|}}}}

\usepackage{hyperref}

\begin{document}

\title{Deep chroma compression of tone-mapped images}

\author{
    Xenios Milidonis\orcidicon{0000-0002-5446-9459},
    Francesco Banterle\orcidicon{0000-0002-6374-6657},
    Alessandro Artusi\orcidicon{0000-0002-4502-663X} 

    \thanks{The research work of Dr. Alessandro Artusi and Dr. Xenios Milidonis has been partially funded by the European Union’s Horizon 2020 Research and Innovation Programme under grant agreement No. 739578 and by the Government of the Republic of Cyprus through the Deputy Ministry of Research, Innovation and Digital Policy.} 

    \thanks{Xenios Milidonis is with the DeepCamera Multidisciplinary Research Group, CYENS Centre of Excellence, Nicosia 1016, Cyprus (email: x.milidonis@cyens.org.cy).}
    \thanks{Francesco Banterle is with the Institute of Information Science and Technologies of the National Research Council of Italy (ISTI-CNR), Pisa 56124, Italy (email: francesco.banterle@isti.cnr.it).}
    \thanks{Alessandro Artusi is with the DeepCamera Multidisciplinary Research Group, CYENS Centre of Excellence, Nicosia 1016, Cyprus (email: a.artusi@cyens.org.cy).}
}

\maketitle

\begin{abstract}
Acquisition of high dynamic range (HDR) images is thriving due to the increasing use of smart devices and the demand for high-quality output. Extensive research has focused on developing methods for reducing the luminance range in HDR images using conventional and deep learning-based tone mapping operators to enable accurate reproduction on conventional 8 and 10-bit digital displays. However, these methods often fail to account for pixels that may lie outside the target display's gamut, resulting in visible chromatic distortions or color clipping artifacts. Previous studies suggested that a gamut management step ensures that all pixels remain within the target gamut. However, such approaches are computationally expensive and cannot be deployed on devices with limited computational resources. We propose a generative adversarial network for fast and reliable chroma compression of HDR tone-mapped images. We design a loss function that considers the hue property of generated images to improve color accuracy, and train the model on an extensive image dataset. Quantitative experiments demonstrate that the proposed model outperforms state-of-the-art image generation and enhancement networks in color accuracy, while a subjective study suggests that the generated images are on par or superior to those produced by conventional chroma compression methods in terms of visual quality. Additionally, the model achieves real-time performance, showing promising results for deployment on devices with limited computational resources.
\end{abstract}

\begin{IEEEkeywords}
High Dynamic Range, Deep Learning, Tone Mapping, Chroma Compression, Generative Adversarial Networks.
\end{IEEEkeywords}

\begin{figure*}
    \centering
    \includegraphics[width=1\textwidth]{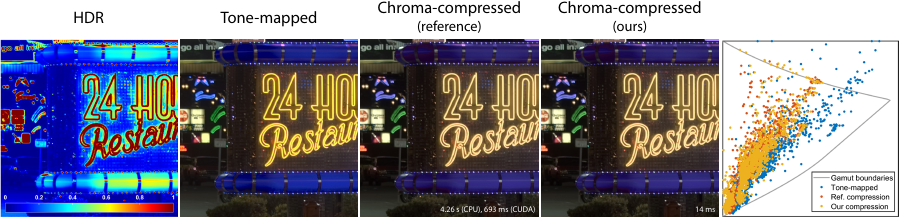}
    \caption{Accurate and efficient chroma compression is important for the reproduction of HDR images on standard displays with limited computational resources. This study proposes a deep learning-based method for chroma compression that is up to 3 orders of magnitude faster than the conventional approach, while generating pixel values that similarly fall within the target display's gamut (here $sRGB$). The scatter plot shows the chroma values ($x$ axis) of pixels within a hue slice ($h=60^{\circ}$) against their lightness values ($y$ axis). Tone mapping cannot compress all pixels within the gamut boundaries, a problem that chroma compression is able to solve.}
    \label{fig-main}
\end{figure*}
\vspace{-2mm}

\section{Introduction}

\IEEEPARstart{H}{igh} dynamic range (HDR) imaging is an important technology targeting the acquisition, display, and storage of high-fidelity images that contain a considerably wider luminance spectrum than conventional images \cite{Banterle:2017}. HDR cameras \cite{Tocci+11}, modern setups \cite{Alghamdi+21}, and smartphones \cite{Hasinoff+16} can acquire a scene with intensities approaching the range perceivable by human vision, reliably capturing shadows and sunlight in the same image. However, due to the limited capacity of traditional monitors to display HDR images, range reduction techniques are employed to generate standard dynamic range (SDR) images that can be accurately reproduced \cite{Artusibook_2016}. These techniques correct primarily the luminance, but subsequent color adjustment may be performed to remove any remaining chromatic distortions or color clipping artifacts (Figure \ref{fig-main}). Most techniques do not consider color correction. The norm in HDR management on SDR display systems is the use of a tone-mapping operator (TMO) to adjust the luminance alone \cite{Ou_2022, Ledda2005}, which can often generate pixel values lying outside the target display's gamut. When SDR images are displayed, such pixels are clipped to the gamut boundaries and appear over-saturated, affecting the overall conveyed color appearance of the original HDR image.

\u{S}ikudov\'a et al. \cite{sikudova2017} were the first to develop a unified tone and chroma management framework that ensures pixels lie within the target gamut, while requiring no calibrated input or tunable parameters. This technique can be used in conjunction with any existing TMO and employs compression algorithms to correct both lightness and chroma channels. Despite its advantages, the technique is computationally expensive, hampering its deployment and use on devices with limited computational resources. Current trends in digital technology are pushing towards higher resolution displays with faster and more efficient processing capabilities, calling for a further step up in computational efficiency. 

Deep learning offers a unique opportunity to tackle this problem as it can replace complex conventional image processing pipelines with simple fine-tuned models that often match their accuracy while significantly decreasing computational costs. In this work, we propose the use of an image-to-image generative model tailored for fast and reliable gamut management of tone-mapped HDR images. Following extensive experiments with state-of-the-art (SOTA) image generation and enhancement models, our approach exploits the seminal generative adversarial networks (GANs) \cite{isola2017image}. We correct undesirable chromatic effects produced by the prior step of tone mapping through careful network optimization, and design a custom loss function to reduce hue distortions. For training, we developed a large dataset augmented by multiple TMOs. The main contributions of this work are:

\begin{itemize}
    \item To the best of our knowledge, the proposed approach is the first deep learning model dedicated to chroma compression and is compatible with all common TMOs, in contrast with existing gamut mapping methods with generally reduced compatibility.
    \item We use a novel loss function that combines the GAN and standard $L1$ losses with a hue-based loss to improve chromatic accuracy.
    \item The model offers significant improvements in computational speed compared to the conventional framework, making it suitable for devices with limited computational resources.
    \item The model is fully automatic and data-driven, requiring no manual input, human expertise, or data other than the tone-mapped image.
    \item The extensive dataset of HDR and SDR images, code and trained weights of the proposed method are publicly available at: \href{https://github.com/DeepCamera/HDR-chroma-compression}{https://github.com/DeepCamera/HDR-chroma-compression}
\end{itemize}

\section{Related work}

\subsection{Chroma compression}

A display system's gamut can be considered as a volume containing the colors that can be reproduced by that display \cite{Artusibook_2016}. If compared with the standard $xy$-chromaticities diagram of the overall colors perceived by the human visual system, a display's gamut is only a subset of it. Tone-mapped images often contain colors lying outside of this gamut, which need to be processed with dedicated chroma correction techniques to be accurately reproduced on displays \cite{Ou_2022}. Chroma correction has been explored for over 2 decades. Initial work attempted to match the viewing conditions between a display and the real-world scene using complex models of the human vision that partially adapted the luminance in the images \cite{pattanaik_1998}. Aky{\"{u}}z and Reinhard \cite{akyuz_2006} later utilized color appearance models for preserving chromatic appearance across various scenes and display environments, but models required manual tuning of multiple parameters which hampered automation and adaptability to variable lighting conditions. Other work tried to develop simple color correction formulas using fewer parameters, focusing either on reducing out-of-gamut colors or enhancing the details in the images but with an impact on the naturalness of the scenes \cite{mantiuk_2009}. Pouli et al. \cite{pouli_2013} presented a parameter-free method for chroma compression with improved visual quality and adaptability to various scenes and TMOs. The same group later extended the algorithm by adding a gamut correction step to further reduce hue shifts and luminance distortion \cite{artusi_2018}. 

In parallel, \u{S}ikudov\'a et al. \cite{sikudova2017} developed a unified framework combining color correction with a gamut management process that reliably handles out-of-gamut pixels. This method first compresses the luminance of an HDR image using any available TMO, corrects the chroma using a novel hue-specific approach that models the target gamut, and finally both luminance and chroma are processed to ensure all pixels remain within the gamut boundaries. The authors showed that this approach is superior to aforementioned methods in handling all pixels while avoiding desaturated or over-saturated results and preserving the details and naturalness of the scene. For these reasons, this method is widely considered among the SOTA in chroma compression, despite its development in the previous decade. However, its algorithmic complexity and reliance on both the HDR and tone-mapped SDR images for calculations lead to a substantial computational burden impacting its deployment potential.

\subsection{Deep learning in HDR imaging}

Deep learning methods for processing HDR content have gained considerable attention in recent years, due to the significant improvement in computational speed they offer. Various neural network architectures can learn contextual information in the images and produce realistic outputs in under- or over-saturated regions, something that conventional processing pipelines are lacking \cite{wang2021deep}. Nevertheless, the vast majority of methods target tone mapping or inverse tone mapping (the reconstruction of HDR from SDR data) \cite{Ou_2022, wang2021deep}, and no published work has tackled the important problem of chroma compression for accurate image visualization.

A considerable effort was put into the reconstruction of HDR images from multi-exposure SDR images \cite{Kalantari+17} or even from a single-exposure image \cite{Eilertsen2017, marnerides2019expandnet} using convolutional neural networks (CNN). Alternatively, Liu et al. \cite{Liu2020} modeled the conventional HDR-to-SDR pipeline and developed a sequence of CNNs to learn the inverse process of HDR reconstruction. One of the scarce models that targeted HDR color enhancement using CNNs is the popular HDRNet \cite{gharbi2017deep}, a multi-scale encoder-decoder that learns local affine transformations from a low-resolution version of the input image without the need to perform full image generation.

GANs have also been used for HDR image processing due to their ability to learn the true data distribution \cite{Goodfellow}. The highly successful implementations for paired and unpaired image-to-image translation, Pix2pix \cite{isola2017image} and CycleGAN \cite{CycleGAN2017} respectively, have recently been adapted for unsupervised tasks in HDR imaging other than color enhancement \cite{Patel2018, Yang2020, Huang2021, Niu2021}. The adversarial loss in GANs plays an important role in the production of realistic outputs that follow the input data distribution, while the addition of a cycle consistency loss in the case of CycleGAN \cite{CycleGAN2017} allows the preservation of the semantic content present in the original images, which could represent the details and intensity levels in the context of HDR imaging. Wang et al. \cite{wang2018pix2pixHD} later developed a multi-scale extension of Pix2pix suitable for high-resolution images. However, to our knowledge, no GAN has yet focused on chroma compression.

Driven by the inability of the standard $L1$ and $L2$ metrics to account for high-level image features and their typical use on non-perceptually uniform color spaces, studies in HDR imaging often use perceptual losses to ensure the production of images with improved visual quality \cite{moriwaki2018hybrid, Santos_2020, wang2021deep}. However, while realistic and visually pleasing images are generally desirable, perceptually-guided image generation risks producing out-of-gamut pixels similar to tone mapping techniques. Alternatively, others added a cosine similarity term in the $L1$ loss to improve color stability during SDR-to-HDR translation, measured in the input $RGB$ color space \cite{marnerides2019expandnet}.

\section{Method}

\subsection{Chroma compression framework}

To train the proposed deep learning-based chroma compression method, we considered the unified gamut management framework by \u{S}ikudov\'a et al. \cite{sikudova2017} (Figure \ref{fig-studyprocess}). In this framework, luminance and color are separated according to the standard $CIELCh$ color space assuming a specific target gamut; here, we used the $sRGB$ with $D65$ white point, the boundaries of which take the form of a triangular cusp for each hue value in $CIELCh$ coordinates (Figure \ref{fig-main}). The chroma is compressed by linearly aligning the input and target chroma cusps while maintaining lightness and hue. Fine details in the image are preserved by splitting each hue slice into base and details layers, of which only the base is compressed. Finally, the remaining pixels lying outside the target gamut are clipped along both luminance and chroma channels. The framework was implemented in MATLAB 2022b.

\subsection{Dataset}

We created a dataset covering a diverse range of scenes and lighting conditions. The dataset contains 1467 HDR images at a resolution of 512$\times$512 pixels, captured by the authors or collected from different publicly available HDR repositories \cite{StuttgartHDR, NemotoEPFL, FairchildHDRSurvey, StanfordHDR, UBC_HDR, Hold-Geoffroy+19, Gardner+2017, Akyuz+07, HDRLabs+2013}. We carefully inspected images to exclude identical scenes or images with very low native resolution. From HDR videos, we selected a frame every 3 seconds to avoid duplicating the scene. Furthermore, we removed black and white borders from panoramic or other images.

\begin{figure}
    \centering
    \includegraphics[width=0.47\textwidth]{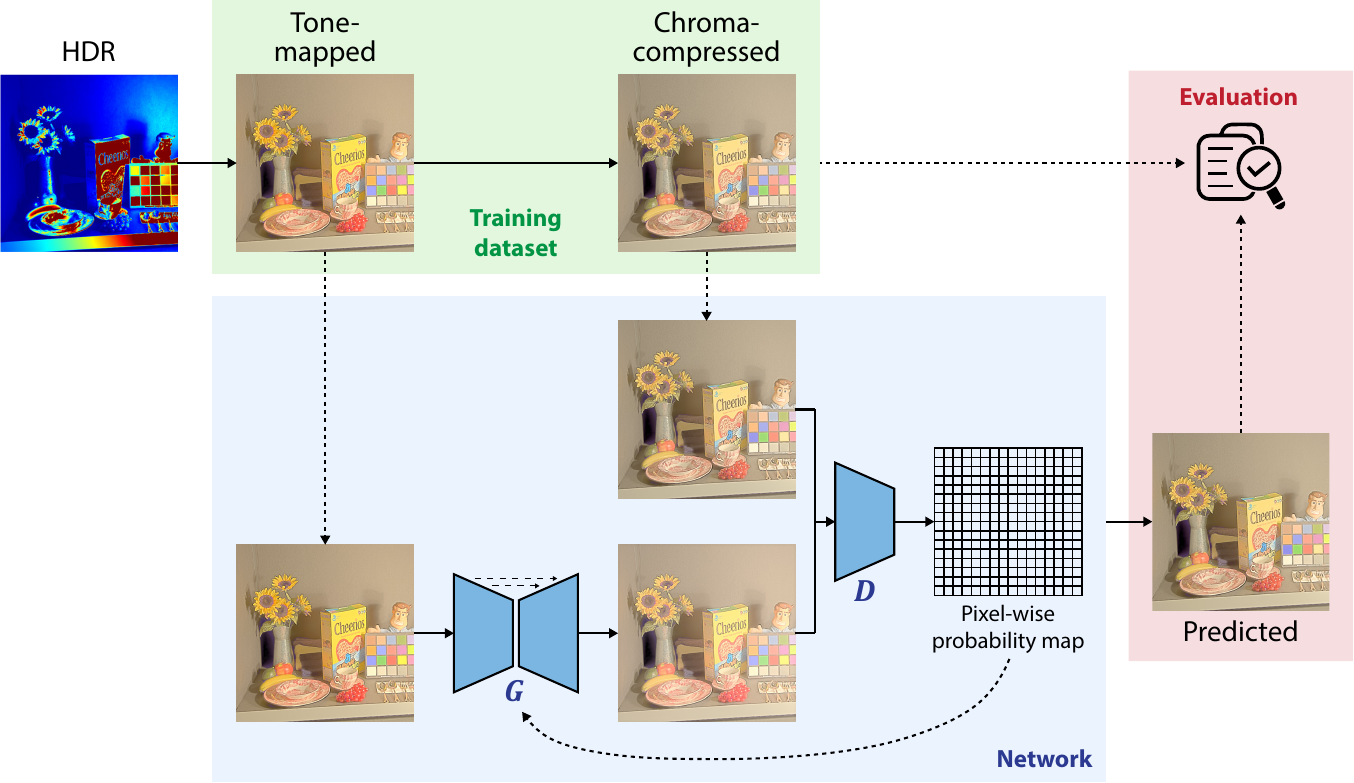}
    \caption{Overview of method and evaluation. HDR images are tone-mapped and then chroma-compressed into SDR images. Our generative adversarial network learns the same process, using a U-Net generator, $G$, and a pixel-based discriminator, $D$. The final images predicted by the trained network are compared against the reference chroma-compressed images for evaluation.}
    \label{fig-studyprocess}
\end{figure}

Rather than using a single TMO to generate the reference SDR images, we used a large number of existing TMOs to augment the dataset and generalize the model (see Table \ref{tab3} for the full list). Tone mapping was performed using a MATLAB toolbox \cite{Banterle:2017}. Images were then chroma-compressed with the chroma compression framework \cite{sikudova2017}.
The final dataset contained a total of 42543 pairs of reference tone-mapped and chroma-compressed images (Figure \ref{fig-studyprocess}), and was split into training, validation, and test subsets at an 8:1:1 ratio (34017, 4263, and 4263 images respectively). TMO versions of the same image were not shared between subsets, as previously suggested \cite{banterle_2023}.

\subsection{Model implementation and training}
\label{model}

We propose the use of a GAN architecture, which has proven effective for an extensive range of image-to-image translation applications \cite{isola2017image}. We have extensively evaluated the performance of this and alternative base models (compared below) and found that it performs more accurately for this task following architecture and hyperparameter optimization. The model consists of a generator and a discriminator, which compete in an adversarial fashion to produce an image and then determine whether it is fake or real, respectively, driving the optimization of the model weights (Figure \ref{fig-studyprocess}). In \cite{isola2017image}, authors used (i) a conditional GAN objective that the generator tries to minimize against an adversarial discriminator that tries to maximize it, and (ii) a generator $L1$ loss to ensure generated images are visually close to the reference images. In this study, we chose a least-squares GAN (LSGAN) objective due to its increased stability during the learning process, leading to the production of higher-quality images compared to standard GANs \cite{mao2017squares}. The GAN loss is the mean squared error (MSE). The model's generator uses a U-Net architecture and a pixel-based rather than a more common patch-based discriminator to reduce tiling artifacts while keeping the details unaffected. 128 generator and 64 discriminator filters are used in the last and first convolutional layers respectively. 

For training, we use the Adam solver with a momentum term of 0.5 and an initial learning rate of $10^{-4}$, with a linear policy and 50 decay iterations. Instance normalization is performed to improve image quality and due to its compatibility with the U-Net architecture \cite{ulyanov2017instance}. Orthogonal weight initialization is used to speed up the convergence. No data pre-processing was applied so that the images were loaded and used for training at their native resolution.

\subsection{Hue-based loss function}

$L1$ and $L2$ loss functions are based on the absolute difference between the reference and predicted image pixel values, and have proven critical in image generation and restoration tasks \cite{wang2021deep}. However, since these losses are traditionally used on non-perceptually uniform color spaces, they consider all pixels equally and do not give weight to regions that are perceptually more important. On the other hand, perceptual metrics alone or in combination with $L1$ and $L2$ were shown to improve the visual quality of images \cite{Zhao2015}, but are less useful in tasks requiring chromatic consistency (i.e. constraining rather than expanding the images' gamut), as the one concerned in this study. Since the task of color correction focuses primarily on the image's hue, we propose the use of an improved loss function that shifts the attention to the hue channel to improve the chromatic accuracy of the result. In the context of conditional GANs, the proposed hue loss can be expressed as

\begin{equation}
\label{eqn1}
\mathcal{L}_{H}(G)=\|I_H-\tilde{I}_H\|_2^2 \text{,}
\end{equation}

\noindent where $I_H$ is the hue channel of the true (reference) image, and $\tilde{I}_H$ is the hue channel of the generated image. Images are normalized between $[0,1]$ before being converted to the $CIELCh$ color space, from which the hue channel is obtained. The loss is combined with the conditional GAN objective to capture global image statistics and perceptual realism to enhance the generalization capabilities of the network, as well as the standard $L1$ loss to improve image production accuracy at the local level. The final objective is given below:

\begin{equation}
\label{eqn2}
G^*=\arg\min_{G}\max_{D}\mathcal{L}_{cGAN}(G,D)+\lambda_1\mathcal{L}_{L1}(G)+\lambda_H\mathcal{L}_{H}(G)
\end{equation}

In equation \eqref{eqn1}, the hue loss is measured in an $L2$ sense due to a small improvement in performance when used in this context, compared to the $L1$ norm. Weights $\lambda_1$ and $\lambda_H$ were fine-tuned independently and in combination; the optimal values were found to be 100 and 10 respectively, which encourage the model to favor pixel-wise accuracy over its generative ability.

\section{Evaluation}
\label{eval}

\noindent In all experiments described below, unless otherwise stated, the same test dataset was used for an unbiased evaluation. Reference and predicted images were converted to double-precision floating-point format and normalized between $[0,1]$. The performance of the models was evaluated using standard metrics (peak signal-to-noise ratio (PSNR), and structural similarity index (SSIM)) \cite{Ou_2022}, metrics that focus on color accuracy (mean absolute error of the hue channel based on the $CIELCh$ color space ($\text{MAE}_H$), and root mean squared error of the color channels in the $CIELab$ color space ($\Delta\text{E}_{Lab}$)) \cite{gaurav2003}, and a recent color-aware visual quality metric (ColorVideoVDP (CVVDP)) \cite{ColorVideoVDP}. CVVDP has just-objectionable-difference (JOD) units with a maximum possible value of 10 indicating no difference.
Model performance at the local level was evaluated using the CVVDP distortion map, which overlays a heatmap with pixel-wise JOD differences between 0 and 1 over a grayscale version of the image \cite{ColorVideoVDP}.
Models were trained and tested on a GPU cluster with 8 nodes, each with 8 NVIDIA RTX A5000 (24 GB on-board GPU memory), running Rocky Linux 8.5 and using Python 3.10 and PyTorch 1.13. Test metrics and processing times for the various models are reported using the mean and were compared using one-way analysis of variance. Statistical significance was set at $p<0.05$.

\begin{figure*}
    \centering
    \includegraphics[width=1\textwidth]{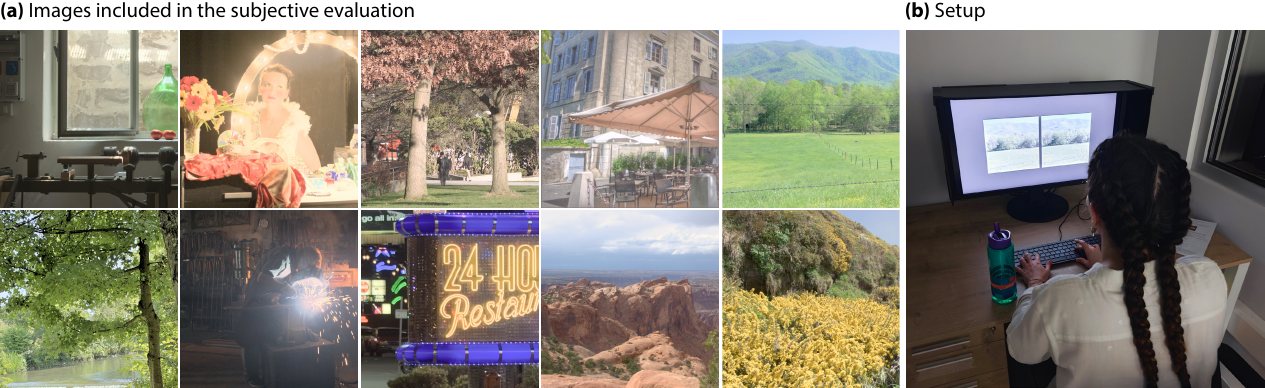}
    \caption{(a) Selected images for the subjective evaluation. (b) The experimental setup in a room with dim lights on for visualization purposes.}
    \label{fig-subjective}
\end{figure*}

\subsection{Comparison against SOTA models}
\label{models}

There are no published neural networks for chroma compression of SDR images. Therefore, the performance of the proposed model was compared against well-established image-to-image translation models, including three GANs and an HDR-focused transformation model: Pix2pix \cite{isola2017image}, CycleGAN \cite{CycleGAN2017}, Pix2pixHD \cite{wang2018pix2pixHD}, and HDRNet \cite{gharbi2017deep}. All models were trained from scratch and tested using the same SDR images as our method, and using the parameters described in the corresponding papers. Training was performed for 200 epochs for all models; HDRNet converges faster than the GANs, but it was found that the same number of epochs provides a performance improvement over fewer epochs. 

\subsection{Component contribution}

In this experiment, the proposed model was trained sequentially by replacing critical components with other common techniques to determine their impact on performance. A model trained without an adversarial loss ($L1$ only) to focus exclusively on pixel-wise accuracy was also considered.

\subsection{Leave-one-out TMO evaluation}

To evaluate the generalizability of the model, leave-one-out experiments were performed by training the model on SDR images tone-mapped with all except one TMO. The model was then tested on images tone-mapped with the excluded TMO. This was repeated for five randomly selected TMOs.

\subsection{Subjective evaluation}

To evaluate the perceived visual quality of the images generated by the proposed method compared to dedicated conventional chroma-compression methods \cite{mantiuk_2009, pouli_2013, sikudova2017, artusi_2018}, we designed a 2-alternative forced-choice experiment (2AFC). Ten images were drawn from the test subset, covering both indoor and outdoor scenes captured in either daylight or nighttime and encompassing a large dynamic range of luminance and chromaticity (Figure \ref{fig-subjective} (a)). The study compared the images produced by each method in pairs shown in successive trials (a total of 100 pairs covering all possible combinations between ours and four existing methods). Each pair of images was displayed side-by-side over a dark grey background (Figure \ref{fig-subjective} (b)), and between trials, a light grey background was shown for 3 seconds. Both the order of the two images in each pair and the order of trials were randomized. In each trial, participants were asked to select the image that they preferred based on perceived realism/naturalness and overall visual quality. Before the experiment, participants were given written instructions and informed consent was obtained. The experiment was performed on a 4K HDR hardware calibration LCD monitor (Eizo ColorEdge CG319X), located in a dark room. The monitor was set to the standard SDR mode, the $sRGB$ color space with $D65$ white point, and a peak luminance of 100 $cd/m^2$.
Votes for each method were assessed using paired comparison standard analysis \cite{David1988}, which has previously been used in computer graphics and is equivalent to analysis of variance \cite{Ledda2005, Banterle2009}. Methods were grouped using the separating $R=90$ value at significance level $\alpha=0.05$.

\subsection{Computational performance}

The inference time during testing of the proposed model was measured to evaluate its computational performance in comparison to the reference conventional framework \cite{sikudova2017}. A batch size of 1 was used to process each image in series with a single GPU. For the conventional approach, both CPU and CUDA implementations of the chroma compression process were considered.

\begin{figure*}
    \centering
    \includegraphics[width=1\textwidth]{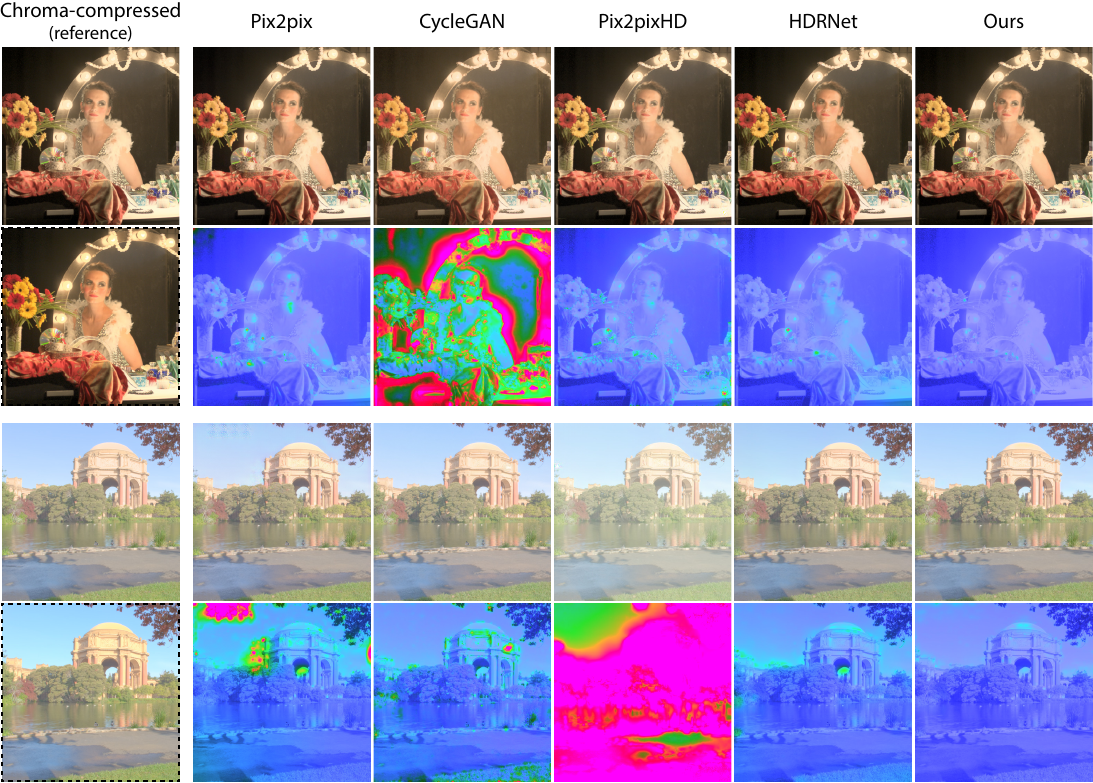}
    \caption{An example image tone-mapped (dashed frame) and subsequently chroma-compressed by the reference method, various SOTA models for image restoration, and our method. Corresponding CVVDP distortion maps are also shown (blue indicates smaller distortion and red/pink indicates larger distortion).}
    \label{fig-2imagecomparison}
\end{figure*}

\section{Results}

\begin{figure}
    \centering
    \includegraphics[width=0.48\textwidth]{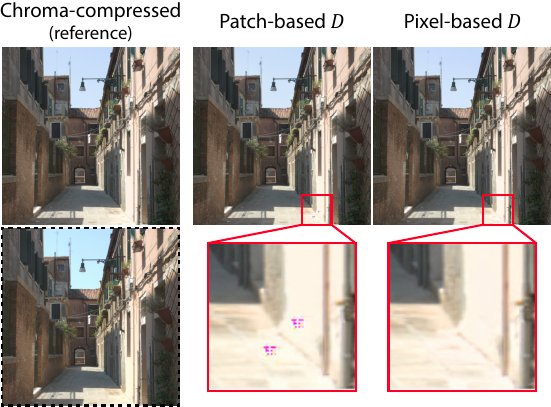}
    \caption{The effect of the size of the discriminator's receptive field. The 70$\times$70 patch-based discriminator occasionally generates tiling artifacts that the pixel-based discriminator can suppress. The original tone-mapped image has a dashed frame.}
    \label{fig-pixelpatch}
\end{figure}

\noindent Figure \ref{fig-2imagecomparison} and Table \ref{tab1} compare the performance of the proposed model against common image generation and dedicated HDR image enhancement models. Our model significantly outperforms considered methods based on all metrics in Table \ref{tab1} ($p<0.001$). Amongst the previous models, HDRNet was found to have the next best performance in chroma compression, followed by Pix2pix. It should be emphasized that efforts to improve the performance of these models suggested that an optimized GAN architecture leads to higher overall pixel-wise accuracy; this could be because GANs target both global patterns and fine-grained local features more effectively than traditional CNN-based methods, making them more compatible with global and local TMOs \cite{gharbi2017deep}. 
At the local level, CVVDP distortion maps in Figure \ref{fig-2imagecomparison} demonstrate the performance of all models in the production of images with accurate colors. The standard Pix2pix generally performs well, but often generates tiling artifacts that can be missed by qualitative evaluation (Figure \ref{fig-pixelpatch}). Despite being an unsupervised model, CycleGAN can preserve the clarity and colorfulness of the images. However, our results indicate that CycleGAN and Pix2pixHD lack consistency in the production of color-accurate results (Figure \ref{fig-2imagecomparison}). 
Our simple GAN architecture significantly improves pixel-wise accuracy with considerably fewer chromatic distortions, while being effective in suppressing any artifacts generated by other models.

\begin{table}
    \caption{Model comparison}
    \vspace{-2mm}
    \label{tab1} 
    \fontsize{8}{8}\selectfont
    \setlength{\tabcolsep}{2pt}
    \renewcommand{\arraystretch}{1.25}
    \begin{minipage}{\columnwidth}
    \begin{center}
    \begin{tabular}{ M{1.5cm} | M{0.9cm} | M{0.8cm} | M{1.0cm} | M{1.0cm} | M{1.1cm} }
    \hline
    Model & PSNR (dB) $\uparrow$ & SSIM $\uparrow$ & $\text{MAE}_H$ $\downarrow$ & $\Delta\text{E}_{Lab}$ $\downarrow$ & CVVDP (JOD) $\uparrow$ \\
    \hline
    Pix2pix             & 35.7 & 0.941 & 0.463 & 2.72 & 9.63\\
    CycleGAN            & 28.6 & 0.921 & 0.489 & 4.00 & 9.28\\
    Pix2pixHD           & 27.5 & 0.898 & 0.533 & 4.61 & 9.14\\
    HDRNet              & 40.6 & 0.982 & 0.250 & 1.60 & 9.82\\
    Ours                & \textbf{43.6} & \textbf{0.993} & \textbf{0.168} & \textbf{1.06} & \textbf{9.91} \\
    \hline
    \end{tabular}
    \end{center}
    \end{minipage}
\end{table}

Various components contribute to the increase in model performance. As shown in Table \ref{tab2}, the pixel-based discriminator increases PSNR by 1.7 dB compared to a $70\times70$ PatchGAN discriminator, instance normalization increases SSIM by 0.005 compared to batch normalization, and orthogonal weight initialization improves $\text{MAE}_H$ by 0.024 compared to normal initialization. Additionally, the hue-based loss function improves $\text{MAE}_H$ by 0.006 compared to the standard loss ($GAN + L1$). A vanilla loss function without an adversarial or hue-based component ($L1$) provides similar performance across all metrics, but is still inferior overall to our full hue-based loss.

\begin{figure}
    \centering
    \includegraphics[width=0.48\textwidth]{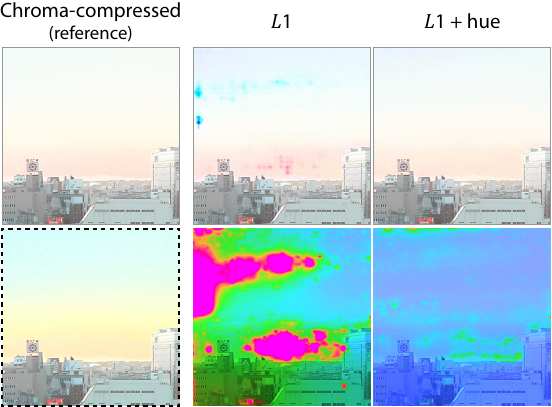}
    \caption{The effect of the hue-based loss function on model performance. Adding a hue loss in the final objective can improve color accuracy in challenging high-hue or saturated regions. Corresponding CVVDP distortion maps are also shown (blue indicates smaller distortion and red/pink indicates larger distortion). The original tone-mapped image has a dashed frame.}
    \label{fig-hueloss}
\end{figure}

Figure \ref{fig-pixelpatch} shows the impact that the size of the discriminator's receptive field has on the generated images. A standard patch-based discriminator \cite{isola2017image} often generates tiling artifacts in the images, particularly in bright areas and usually small in size. These artifacts are removed by the use of a discriminator evaluating all individual pixels (i.e., $1\times1$ patches), without affecting the colorfulness or spatial sharpness of the results. Figure \ref{fig-hueloss} demonstrates the effect of the hue-based loss function. Despite not having an effect on image sharpness, generated images tend to have improved chromatic accuracy in challenging regions with high hue or saturation values.

\begin{table}
    \caption{Contribution of Model Components. In Each Row, we Replaced a Component of Our Model, as Described in Section \ref{model}, With Alternative Common Techniques. All Other Model Parameters in Each Row Remain Fixed}
    \vspace{-2mm}
    \label{tab2} 
    \fontsize{8}{8}\selectfont
    \setlength{\tabcolsep}{2pt}
    \renewcommand{\arraystretch}{1.25}
    \begin{minipage}{\columnwidth}
    \begin{center}
    \begin{tabular}{ M{2.2cm} | M{0.9cm} | M{0.8cm} | M{1.0cm} | M{1.0cm} | M{1.1cm} }
    \hline
    Component & PSNR (dB) $\uparrow$ & SSIM $\uparrow$ & $\text{MAE}_H$ $\downarrow$ & $\Delta\text{E}_{Lab}$ $\downarrow$ & CVVDP (JOD) $\uparrow$ \\
    \hline
    Patch-based $D$                 & 41.9 & 0.991 & 0.175 & 1.23 & 9.86\\
    Image-based $D$                 & 43.1 & 0.992 & 0.175 & 1.10 & 9.90\\
    64 $G$ filters                  & 42.6 & 0.991 & 0.190 & 1.18 & 9.89\\
    Batch norm.                     & 42.4 & 0.988 & 0.183 & 1.21 & 9.89\\
    Normal init.                    & 43.0 & 0.990 & 0.192 & 1.16 & 9.89\\
    Cross-entropy obj.              & 43.5 & 0.992 & 0.174 & 1.08 & \textbf{9.91}\\
    $(GAN + L1)$ loss               & 43.5 & 0.992 & 0.174 & 1.07 & 9.90\\
    $L1$ loss                       & \textbf{43.6} & 0.992 & 0.170 & \textbf{1.06} & \textbf{9.91}\\
    Ours                            & \textbf{43.6} & \textbf{0.993} & \textbf{0.168} & \textbf{1.06} & \textbf{9.91} \\
    \hline
    \end{tabular}
    \end{center}
    \end{minipage}
\end{table}

Table \ref{tab3} summarizes the performance of the proposed model on images tone-mapped by individual TMOs. The proposed method handles images tone-mapped by various TMOs with high accuracy, with PSNR ranging between 40.2 dB and 47.5 dB, SSIM between 0.983 and 0.997, $\text{MAE}_H$ between 0.108 and 0.261, $\Delta\text{E}_{Lab}$ between 0.65 and 1.55, and CVVDP between 9.81 JOD and 9.95 JOD. 
The results of leave-one-out experiments are given at the bottom of Table \ref{tab3}. All performance metrics suggest that each model's accuracy is on par with the model trained on images from all TMOs and generalizes well to the unseen TMOs.

\begin{table}
    \caption{Performance for Individual TMOs. The YP Methods Use Yee and Pattanaik's Luminance Adaptation \cite{YeePattanaik}. Five Methods Using Basic Luminance Operations Were Also Considered (*). Below, the Metrics for the Five Leave-one-out TMO Experiments are Provided (Values For the Same TMOs May Differ Also Due to the Different Size of Test Subsets).}
    \vspace{-2mm}
    \label{tab3}
    \fontsize{8}{8}\selectfont
    \setlength{\tabcolsep}{2pt}
    \renewcommand{\arraystretch}{1.25}
    \begin{minipage}{\columnwidth}
    \begin{center}
    \begin{tabular}{ M{2.2cm} | M{0.9cm} | M{0.8cm} | M{1.0cm} | M{1.0cm} | M{1.1cm} }
    \hline
    TMO & PSNR (dB) $\uparrow$ & SSIM $\uparrow$ & $\text{MAE}_H$ $\downarrow$ & $\Delta\text{E}_{Lab}$ $\downarrow$ & CVVDP (JOD) $\uparrow$ \\
    \hline
    Ashikhmin \cite{AshikhminTMO}   & 43.5 & 0.995 & 0.142 & 1.02 & 9.91\\
    Banterle \cite{BanterleTMO}     & 42.9 & 0.994 & 0.143 & 1.11 & 9.90\\
    Bruce \cite{BruceTMO}           & 42.1 & 0.993 & 0.155 & 1.23 & 9.89\\
    Chiu \cite{ChiuTMO}             & 42.6 & 0.986 & 0.184 & 1.29 & 9.88\\
    Drago \cite{DragoTMO}           & 42.6 & 0.992 & 0.162 & 1.17 & 9.89\\
    Durand \cite{DurandTMO}         & 41.9 & 0.993 & 0.151 & 1.26 & 9.88\\
    Fattal \cite{FattalTMO}         & 42.3 & 0.992 & 0.154 & 1.19 & 9.89\\
    Ferwerda \cite{FerwerdaTMO}     & 45.5 & 0.993 & 0.190 & 0.82 & 9.93\\
    Kim-Kautz \cite{KimKautzTMO}    & 44.3 & 0.995 & 0.145 & 0.95 & 9.93\\
    Krawczyk \cite{KrawczykTMO}     & 43.4 & 0.992 & 0.171 & 1.05 & 9.90\\
    Kuang \cite{KuangTMO}           & 45.3 & 0.995 & 0.108 & 0.83 & 9.95\\
    Li \cite{LiTMO}                 & 43.4 & 0.997 & 0.133 & 1.01 & 9.92\\
    Lischinski \cite{LischinskiTMO} & 44.7 & 0.995 & 0.130 & 0.92 & 9.93\\
    Mertens \cite{MertensTMO}       & 47.5 & 0.994 & 0.217 & 0.65 & 9.95\\
    Raman \cite{RamanTMO}           & 41.1 & 0.983 & 0.261 & 1.40 & 9.88\\
    Reinhard \cite{ReinhardTMO}     & 44.0 & 0.995 & 0.134 & 1.00 & 9.92\\
    Reinh.-Devlin \cite{ReinhardDevlinTMO} & 41.6 & 0.990 & 0.184 & 1.34 & 9.88\\
    Schlick \cite{SchlickTMO}       & 44.9 & 0.994 & 0.158 & 1.02 & 9.92\\
    Tumblin \cite{TumblinTMO}       & 43.3 & 0.992 & 0.159 & 1.12 & 9.90\\
    Tumblin (YP)                    & 41.6 & 0.990 & 0.190 & 1.33 & 9.88\\
    Van Hateren \cite{VanHaterenTMO} & 44.0 & 0.990 & 0.176 & 1.07 & 9.91\\
    Ward \cite{WardTMO}             & 45.0 & 0.996 & 0.162 & 0.84 & 9.93\\
    Ward global \cite{WardGlobalTMO} & 44.6 & 0.993 & 0.175 & 0.93 & 9.91\\
    Ward global (YP)                & 44.4 & 0.993 & 0.161 & 0.95 & 9.92\\
    Best exposure*                  & 44.9 & 0.994 & 0.153 & 0.90 & 9.92\\
    Exponential*                    & 40.2 & 0.985 & 0.226 & 1.55 & 9.81\\
    Logarithmic*                    & 44.2 & 0.992 & 0.175 & 1.00 & 9.91\\
    Normalize*                      & 46.0 & 0.994 & 0.221 & 0.74 & 9.94\\
    Photographic*                   & 43.6 & 0.994 & 0.142 & 1.05 & 9.91\\
    \hline
    Chiu                            & 40.7 & 0.990 & 0.150 & 1.08 & 9.91\\
    Krawczyk                        & 41.1 & 0.994 & 0.142 & 0.87 & 9.92\\
    Reinhard                        & 43.9 & 0.998 & 0.102 & 0.68 & 9.95\\
    Van Hateren                     & 43.5 & 0.989 & 0.167 & 0.85 & 9.93\\
    Logarithmic*                    & 44.0 & 0.995 & 0.146 & 0.72 & 9.95\\
    \hline
    \end{tabular}
    \end{center}
    \end{minipage}
\end{table}

\begin{figure}
    \centering
    \includegraphics[width=0.48\textwidth]{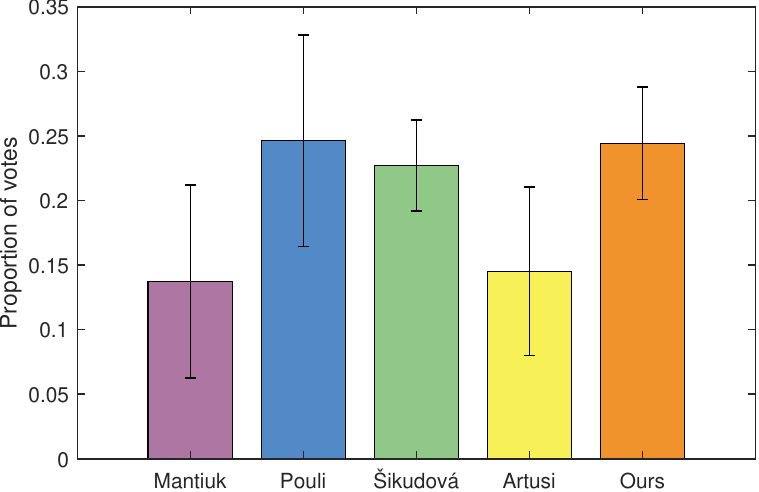}
    \caption{Results of the subjective evaluation. The graph shows the proportion of times each method was selected by participants. Error bars indicate the standard deviation.}
    \label{fig-votes}
\end{figure}

In the subjective study, we recruited 21 participants (mean age 33 [24 – 58] years, 15 men and 6 women), all having a normal or corrected-to-normal vision and no color blindness.
Table \ref{tab4} shows the results of paired comparison standard analysis \cite{David1988}. The coefficients of consistency and agreement are 0.810 (possible range [0, 1]) and 0.093 (possible range [-1, 1]) respectively, suggesting that there is high consistency overall in voting and a positive agreement amongst voters  \cite{David1988}). Participants agree on voting at a statistically significant level based on a $X^2$ of 204.76 (critical value at $\alpha=0.05$ is 18.31), and differences in scores were significant based on a $D_n$ of 197.93 (critical value at $\alpha=0.05$ is 9.45). The analysis revealed that our neural network is equivalent to Pouli's and \u{S}ikudov\'a's analytic methods in producing visually pleasing images, and was selected significantly more often than Artusi's and Mantiuk's methods (Figure \ref{fig-votes}). Pouli's method was chosen less consistently than our and \u{S}ikudov\'a's methods (standard deviation is 0.82 versus 0.44 and 0.35 respectively). 

\begin{figure}
    \centering
    \includegraphics[width=0.48\textwidth]{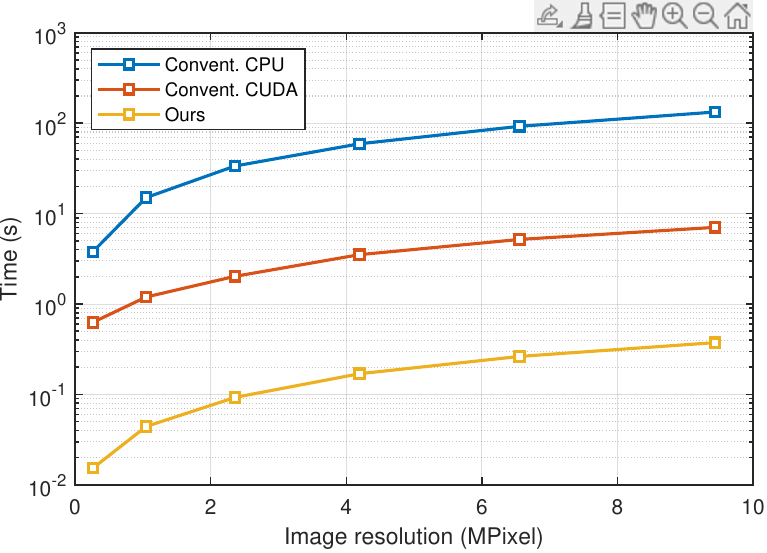}
    \caption{Mean inference time per image for the conventional chroma compression algorithm (CPU and CUDA versions) and our approach.}
    \label{fig-proctime}
\end{figure}

Computational performance at inference time is significantly better for the proposed model compared to the conventional chroma compression framework at all tested resolutions ($p<0.001$; Figure \ref{fig-proctime}). For example, for the smallest resolution of 512$\times$512 (0.26 MPixel), our approach takes 15 $ms$ per image, compared to 627 $ms$ for the conventional CUDA and 3.8 $s$ for the conventional CPU approaches (41-fold and 247-fold improvement respectively). Real-time compression is maintained up to 1 MPixel with 25 frames per second, bringing the performance of chroma compression at the level of optimized tone mapping techniques \cite{Ou_2022}. At the highest tested resolution of 9.4 MPixel, the model can process images significantly more efficiently (375 $ms$, compared to 7.1 $s$ for the conventional CUDA and 133.0 $s$ for conventional CPU approaches).

\begin{table}
    \caption{Analysis of the Subjective Evaluation. The Top Row Provides the Coefficients of Consistency and Agreement, as well as the $X^2$ Value for the Coefficients of Agreement and the $D_n$ Value for the Test of Equality to Determine the Significance of our Data. The Bottom Row Reports the Ranking of the Methods with Total Votes in Parentheses. We Group Together Methods that are Statistically the Same Using the Critical Value $R=90$ at Significance Level $\alpha=0.05$. A Indicates Artusi; M, Mantiuk; O, Our Model; P, Pouli; and S, \u{S}ikudov\'a}
    \vspace{-2mm}
    \label{tab4} 
    \fontsize{8}{8}\selectfont
    \setlength{\tabcolsep}{2pt}
    \renewcommand{\arraystretch}{1.25}
    \begin{minipage}{\columnwidth}
    \begin{center}
    \begin{tabular}{ M{1.4cm} | M{1.4cm} | M{1.4cm} | M{0.9cm} | M{0.9cm} }
    \hline
         & Coeff. Consistency & Coeff. Agreement & $X^2$ & $D_n$\\
    \hline
     Overall & 0.810 & 0.093 & 204.76 & 197.93 \\
    \hline
    \hline
          & \multicolumn{4}{c}{Groups}\\
    \mymkw{Overall} & \multicolumn{4}{c}{\mymkb{M(288) A(305)}} \mymkr{S(477)  O(513) P(517)}\\
    \hline
    \end{tabular}
    \end{center}
    \end{minipage}
\end{table}

\section{Conclusions}

In this work, we proposed a method for fast, accurate, and generalizable chroma compression of tone-mapped images. To the best of our knowledge, this is the first deep learning approach performing chroma compression, offering excellent performance in terms of pixel-wise accuracy, color accuracy, and imaging artifacts. Importantly, the model's performance based on PSNR and SSIM is superior to that reported for a variety of other tasks in HDR imaging \cite{wang2021deep, gharbi2017deep}. 
Despite being of secondary importance, the proposed method also matches \cite{pouli_2013, sikudova2017} or surpasses \cite{mantiuk_2009, artusi_2018} the visual quality of the results of conventional SOTA frameworks for chroma compression. The reference framework by \u{S}ikudov\'a et al. \cite{sikudova2017} was chosen as the foundation due to its superiority over competing methods, which can generate similarly visually pleasing results (e.g. \cite{pouli_2013}) but are more susceptible to hue distortion artifacts and out-of-gamut pixels, as previously demonstrated \cite{sikudova2017}. Therefore, our model, trained to be a significantly faster alternative to the conventional framework, shows great promise for applications where efficient and accurate color reproduction is of paramount importance. Our work offers a simple solution that could be implemented on devices with limited computational resources and software employing any TMO for handling HDR data, such as smartphones and tablets, reducing chromatic artifacts in captured images and videos. 
Nevertheless, the proposed method is limited to processing images for specific display properties ($sRGB$ gamut), and further work is needed to develop a general-purpose chroma compression approach. Finally, while the proposed method is theoretically adaptable to video processing, future research could incorporate it into a pipeline for real-time HDR video management.

\bibliographystyle{IEEEtran}
\bibliography{bib}

\begin{thebibliography}{10}
\providecommand{\url}[1]{#1}
\csname url@samestyle\endcsname
\providecommand{\newblock}{\relax}
\providecommand{\bibinfo}[2]{#2}
\providecommand{\BIBentrySTDinterwordspacing}{\spaceskip=0pt\relax}
\providecommand{\BIBentryALTinterwordstretchfactor}{4}
\providecommand{\BIBentryALTinterwordspacing}{\spaceskip=\fontdimen2\font plus
\BIBentryALTinterwordstretchfactor\fontdimen3\font minus \fontdimen4\font\relax}
\providecommand{\BIBforeignlanguage}[2]{{%
\expandafter\ifx\csname l@#1\endcsname\relax
\typeout{** WARNING: IEEEtran.bst: No hyphenation pattern has been}%
\typeout{** loaded for the language `#1'. Using the pattern for}%
\typeout{** the default language instead.}%
\else
\language=\csname l@#1\endcsname
\fi
#2}}
\providecommand{\BIBdecl}{\relax}
\BIBdecl

\bibitem{Banterle:2017}
F.~Banterle, A.~Artusi, K.~Debattista, and A.~Chalmers, \emph{Advanced High Dynamic Range Imaging (2nd Edition)}.\hskip 1em plus 0.5em minus 0.4em\relax Natick, MA, USA: AK Peters (CRC Press), July 2017.

\bibitem{Tocci+11}
\BIBentryALTinterwordspacing
M.~D. Tocci, C.~Kiser, N.~Tocci, and P.~Sen, ``A versatile {HDR} video production system,'' \emph{ACM Trans. Graph.}, vol.~30, no.~4, 2011. [Online]. Available: \url{https://doi.org/10.1145/2010324.1964936}
\BIBentrySTDinterwordspacing

\bibitem{Alghamdi+21}
\BIBentryALTinterwordspacing
M.~Alghamdi, Q.~Fu, A.~K. Thabet, and W.~Heidrich, ``Transfer deep learning for reconfigurable snapshot {HDR} imaging using coded masks,'' \emph{Comput. Graph. Forum}, vol.~40, no.~6, pp. 90--103, 2021. [Online]. Available: \url{https://doi.org/10.1111/cgf.14205}
\BIBentrySTDinterwordspacing

\bibitem{Hasinoff+16}
\BIBentryALTinterwordspacing
S.~W. Hasinoff, D.~Sharlet, R.~Geiss, A.~Adams, J.~T. Barron, F.~Kainz, J.~Chen, and M.~Levoy, ``Burst photography for high dynamic range and low-light imaging on mobile cameras,'' \emph{ACM Trans. Graph.}, vol.~35, no.~6, nov 2016. [Online]. Available: \url{https://doi.org/10.1145/2980179.2980254}
\BIBentrySTDinterwordspacing

\bibitem{Artusibook_2016}
A.~Artusi, F.~Banterle, T.~O. Aydn, D.~Panozzo, and O.~Sorkine-Hornung, \emph{Image Content Retargeting: Maintaining Color, Tone, and Spatial Consistency}.\hskip 1em plus 0.5em minus 0.4em\relax USA: A. K. Peters, Ltd., 2016.

\bibitem{Ou_2022}
Y.~Ou, P.~Ambalathankandy, S.~Takamaeda, M.~Motomura, T.~Asai, and M.~Ikebe, ``Real-time tone mapping: A survey and cross-implementation hardware benchmark,'' \emph{IEEE Transactions on Circuits and Systems for Video Technology}, vol.~32, no.~5, pp. 2666--2686, 2022.

\bibitem{Ledda2005}
P.~Ledda, A.~Chalmers, T.~Troscianko, and H.~Seetzen, ``Evaluation of tone mapping operators using a high dynamic range display,'' \emph{{ACM} Trans. Graph.}, vol.~24, no.~3, pp. 640--648, 2005.

\bibitem{sikudova2017}
E.~\u{S}ikudov\'a, T.~Pouli, A.~Artusi, A.~O. Akyüz, F.~Banterle, Z.~M. Mazlumoglu, and E.~Reinhard, ``A gamut-mapping framework for color-accurate reproduction of {HDR} images,'' \emph{IEEE Computer Graphics and Applications}, vol.~36, no.~4, pp. 78--90, 2016.

\bibitem{isola2017image}
P.~Isola, J.-Y. Zhu, T.~Zhou, and A.~A. Efros, ``Image-to-image translation with conditional adversarial networks,'' in \emph{2017 IEEE Conference on Computer Vision and Pattern Recognition (CVPR)}, 2017, pp. 5967--5976.

\bibitem{pattanaik_1998}
\BIBentryALTinterwordspacing
S.~N. Pattanaik, J.~A. Ferwerda, M.~D. Fairchild, and D.~P. Greenberg, ``A multiscale model of adaptation and spatial vision for realistic image display,'' in \emph{Proceedings of the 25th Annual Conference on Computer Graphics and Interactive Techniques}, ser. SIGGRAPH '98.\hskip 1em plus 0.5em minus 0.4em\relax New York, NY, USA: Association for Computing Machinery, 1998, p. 287–298. [Online]. Available: \url{https://doi.org/10.1145/280814.280922}
\BIBentrySTDinterwordspacing

\bibitem{akyuz_2006}
\BIBentryALTinterwordspacing
A.~O. Aky{\"{u}}z and E.~Reinhard, ``Color appearance in high-dynamic-range imaging,'' \emph{Journal of Electronic Imaging}, vol.~15, no.~3, p. 033001, 2006. [Online]. Available: \url{https://doi.org/10.1117/1.2238891}
\BIBentrySTDinterwordspacing

\bibitem{mantiuk_2009}
R.~Mantiuk, R.~Mantiuk, A.~Tomaszewska, and W.~Heidrich, ``Color correction for tone mapping,'' \emph{Computer Graphics Forum}, vol.~28, pp. 193--202, 2009.

\bibitem{pouli_2013}
\BIBentryALTinterwordspacing
T.~Pouli, A.~Artusi, F.~Banterle, A.~O. Aky{\"u}z, H.-P. Seidel, and E.~Reinhard, ``Color correction for tone reproduction,'' in \emph{CIC21: Twenty-first Color and Imaging Conference}, Society for Imaging Science and Technology (IS\&T).\hskip 1em plus 0.5em minus 0.4em\relax Society for Imaging Science and Technology (IS\&T), November 2013, pp. 215--220. [Online]. Available: \url{http://vcg.isti.cnr.it/Publications/2013/PABASR13}
\BIBentrySTDinterwordspacing

\bibitem{artusi_2018}
\BIBentryALTinterwordspacing
A.~Artusi, T.~Pouli, F.~Banterle, and A.~Oğuz Akyüz, ``Automatic saturation correction for dynamic range management algorithms,'' \emph{Signal Processing: Image Communication}, vol.~63, pp. 100--112, 2018. [Online]. Available: \url{https://www.sciencedirect.com/science/article/pii/S0923596518300973}
\BIBentrySTDinterwordspacing

\bibitem{wang2021deep}
L.~Wang and K.-J. Yoon, ``Deep learning for {HDR} imaging: State-of-the-art and future trends,'' \emph{IEEE transactions on pattern analysis and machine intelligence}, vol.~44, no.~12, pp. 8874--8895, 2021.

\bibitem{Kalantari+17}
\BIBentryALTinterwordspacing
N.~K. Kalantari and R.~Ramamoorthi, ``Deep high dynamic range imaging of dynamic scenes,'' \emph{ACM Trans. Graph.}, vol.~36, no.~4, jul 2017. [Online]. Available: \url{https://doi.org/10.1145/3072959.3073609}
\BIBentrySTDinterwordspacing

\bibitem{Eilertsen2017}
\BIBentryALTinterwordspacing
G.~Eilertsen, J.~Kronander, G.~Denes, R.~K. Mantiuk, and J.~Unger, ``{HDR} image reconstruction from a single exposure using deep {CNNs},'' \emph{ACM Trans. Graph.}, vol.~36, no.~6, nov 2017. [Online]. Available: \url{https://doi.org/10.1145/3130800.3130816}
\BIBentrySTDinterwordspacing

\bibitem{marnerides2019expandnet}
D.~Marnerides, T.~Bashford-Rogers, J.~Hatchett, and K.~Debattista, ``Expand{N}et: A deep convolutional neural network for high dynamic range expansion from low dynamic range content,'' in \emph{Computer Graphics Forum}, vol.~37, no.~2.\hskip 1em plus 0.5em minus 0.4em\relax Wiley Online Library, 2018, pp. 37--49.

\bibitem{Liu2020}
Y.-L. Liu, W.-S. Lai, Y.-S. Chen, Y.-L. Kao, M.-H. Yang, Y.-Y. Chuang, and J.-B. Huang, ``Single-image {HDR} reconstruction by learning to reverse the camera pipeline,'' in \emph{2020 IEEE/CVF Conference on Computer Vision and Pattern Recognition (CVPR)}.\hskip 1em plus 0.5em minus 0.4em\relax IEEE, 2020, pp. 1648--1657.

\bibitem{gharbi2017deep}
M.~Gharbi, J.~Chen, J.~T. Barron, S.~W. Hasinoff, and F.~Durand, ``Deep bilateral learning for real-time image enhancement,'' \emph{ACM Transactions on Graphics (TOG)}, vol.~36, no.~4, p. 118, 2017.

\bibitem{Goodfellow}
\BIBentryALTinterwordspacing
I.~Goodfellow, J.~Pouget-Abadie, M.~Mirza, B.~Xu, D.~Warde-Farley, S.~Ozair, A.~Courville, and Y.~Bengio, ``Generative adversarial networks,'' \emph{Commun. ACM}, vol.~63, no.~11, pp. 139--144, October 2020. [Online]. Available: \url{https://doi.org/10.1145/3422622}
\BIBentrySTDinterwordspacing

\bibitem{CycleGAN2017}
J.-Y. Zhu, T.~Park, P.~Isola, and A.~A. Efros, ``Unpaired image-to-image translation using cycle-consistent adversarial networks,'' in \emph{2017 IEEE International Conference on Computer Vision (ICCV)}, 2017, pp. 2242--2251.

\bibitem{Patel2018}
V.~A. Patel, P.~Shah, and S.~Raman, ``A generative adversarial network for tone mapping {HDR} images,'' in \emph{Computer Vision, Pattern Recognition, Image Processing, and Graphics}, R.~Rameshan, C.~Arora, and S.~Dutta~Roy, Eds.\hskip 1em plus 0.5em minus 0.4em\relax Singapore: Springer Singapore, 2018, pp. 220--231.

\bibitem{Yang2020}
Z.~Yang, Y.~Chen, Z.~Le, and Y.~Ma, ``{GANFuse}: a novel multi-exposure image fusion method based on generative adversarial networks,'' \emph{Neural Computing and Applications}, vol.~33, pp. 6133 -- 6145, 2020.

\bibitem{Huang2021}
Y.~Huang, S.~Qiu, C.~Wang, and C.~Li, ``Learning representations for high-dynamic-range image color transfer in a self-supervised way,'' \emph{IEEE Transactions on Multimedia}, vol.~23, pp. 176--188, 2021.

\bibitem{Niu2021}
Y.~Niu, J.~Wu, W.~Liu, W.~Guo, and R.~W.~H. Lau, ``{HDR}-{GAN}: {HDR} image reconstruction from multi-exposed {LDR} images with large motions,'' \emph{{IEEE} Transactions on Image Processing}, vol.~30, pp. 3885--3896, 2021.

\bibitem{wang2018pix2pixHD}
T.-C. Wang, M.-Y. Liu, J.-Y. Zhu, A.~Tao, J.~Kautz, and B.~Catanzaro, ``High-resolution image synthesis and semantic manipulation with conditional {GANs},'' in \emph{2018 IEEE/CVF Conference on Computer Vision and Pattern Recognition}, 2018, pp. 8798--8807.

\bibitem{moriwaki2018hybrid}
K.~Moriwaki, R.~Yoshihashi, R.~Kawakami, S.~You, and T.~Naemura, ``Hybrid loss for learning single-image-based {HDR} reconstruction,'' \emph{arXiv preprint arXiv:1812.07134}, 2018.

\bibitem{Santos_2020}
M.~S. Santos, T.~I. Ren, and N.~K. Kalantari, ``Single image {HDR} reconstruction using a {CNN} with masked features and perceptual loss,'' \emph{{ACM} Transactions on Graphics}, vol.~39, no.~4, August 2020.

\bibitem{StuttgartHDR}
\BIBentryALTinterwordspacing
J.~Froehlich, S.~Grandinetti, B.~Eberhardt, S.~Walter, A.~Schilling, and H.~Brendel, ``Creating cinematic wide gamut {HDR}-video for the evaluation of tone mapping operators and {HDR}-displays,'' in \emph{Digital Photography X}, N.~Sampat, R.~Tezaur, S.~Battiato, and B.~A. Fowler, Eds., vol. 9023, International Society for Optics and Photonics.\hskip 1em plus 0.5em minus 0.4em\relax SPIE, 2014, p. 90230X. [Online]. Available: \url{https://doi.org/10.1117/12.2040003}
\BIBentrySTDinterwordspacing

\bibitem{NemotoEPFL}
\BIBentryALTinterwordspacing
H.~Nemoto, P.~Korshunov, P.~Hanhart, and T.~Ebrahimi, ``Visual attention in {LDR} and {HDR} images,'' in \emph{9th International Workshop on Video Processing and Quality Metrics for Consumer Electronics (VPQM)}.\hskip 1em plus 0.5em minus 0.4em\relax Springer-Verlag, 2015. [Online]. Available: \url{http://infoscience.epfl.ch/record/203873}
\BIBentrySTDinterwordspacing

\bibitem{FairchildHDRSurvey}
M.~D. Fairchild, ``The {HDR} photographic survey,'' in \emph{International Conference on Communications in Computing}.\hskip 1em plus 0.5em minus 0.4em\relax IS\&T Publications, 2007, pp. 233--238.

\bibitem{StanfordHDR}
\BIBentryALTinterwordspacing
F.~Xiao, J.~M. DiCarlo, P.~B. Catrysse, and B.~A. Wandell, ``High dynamic range imaging of natural scenes,'' in \emph{The Tenth Color Imaging Conference: Color Science and Engineering Systems, Technologies, Applications, {CIC} 2002, Scottsdale, Arizona, USA, November 12-15, 2002}.\hskip 1em plus 0.5em minus 0.4em\relax Society for Imaging Science and Technology, 2002, pp. 337--342. [Online]. Available: \url{https://doi.org/10.2352/CIC.2002.10.1.art00062}
\BIBentrySTDinterwordspacing

\bibitem{UBC_HDR}
M.~Azimi, A.~Banitalebi-Dehkordi, Y.~Dong, M.~T. Pourazad, and P.~Nasiopoulos, ``Evaluating the performance of existing full-reference quality metrics on high dynamic range ({HDR}) video content,'' \emph{arXiv preprint arXiv:1803.04815}, 2018.

\bibitem{Hold-Geoffroy+19}
Y.~Hold-Geoffroy, A.~Athawale, and J.-F. Lalonde, ``Deep sky modeling for single image outdoor lighting estimation,'' in \emph{IEEE/CVF Conference on Computer Vision and Pattern Recognition (CVPR)}.\hskip 1em plus 0.5em minus 0.4em\relax IEEE, 2019, pp. 6920--6928.

\bibitem{Gardner+2017}
\BIBentryALTinterwordspacing
M.-A. Gardner, K.~Sunkavalli, E.~Yumer, X.~Shen, E.~Gambaretto, C.~Gagn\'{e}, and J.-F. Lalonde, ``Learning to predict indoor illumination from a single image,'' \emph{ACM Trans. Graph.}, vol.~36, no.~6, November 2017. [Online]. Available: \url{https://doi.org/10.1145/3130800.3130891}
\BIBentrySTDinterwordspacing

\bibitem{Akyuz+07}
\BIBentryALTinterwordspacing
A.~O. Aky{\"{u}}z, R.~W. Fleming, B.~E. Riecke, E.~Reinhard, and H.~H. B{\"{u}}lthoff, ``Do {HDR} displays support {LDR} content?: a psychophysical evaluation,'' \emph{{ACM} Trans. Graph.}, vol.~26, no.~3, p.~38, 2007. [Online]. Available: \url{https://doi.org/10.1145/1276377.1276425}
\BIBentrySTDinterwordspacing

\bibitem{HDRLabs+2013}
\BIBentryALTinterwordspacing
HDRLabs, ``{sIBL} archive free {HDRI} sets for smart image-based lighting,'' 2013. [Online]. Available: \url{https://hdrlabs.com}
\BIBentrySTDinterwordspacing

\bibitem{banterle_2023}
F.~Banterle, A.~Artusi, A.~Moreo, F.~Carrara, and P.~Cignoni, ``Nor-vdpnet++: Real-time no-reference image quality metrics,'' \emph{IEEE Access}, vol.~11, pp. 34\,544--34\,553, 2023.

\bibitem{mao2017squares}
X.~Mao, Q.~Li, H.~Xie, R.~Y. Lau, Z.~Wang, and S.~Paul~Smolley, ``Least squares generative adversarial networks,'' in \emph{Proceedings of the IEEE international conference on computer vision}, 2017, pp. 2794--2802.

\bibitem{ulyanov2017instance}
D.~Ulyanov, A.~Vedaldi, and V.~Lempitsky, ``Instance normalization: The missing ingredient for fast stylization,'' \emph{arXiv preprint arXiv:1607.08022}, 2016.

\bibitem{Zhao2015}
\BIBentryALTinterwordspacing
H.~Zhao, O.~Gallo, I.~Frosio, and J.~Kautz, ``Loss functions for image restoration with neural networks,'' \emph{{IEEE} Trans. Computational Imaging}, vol.~3, no.~1, pp. 47--57, 2017. [Online]. Available: \url{https://doi.org/10.1109/TCI.2016.2644865}
\BIBentrySTDinterwordspacing

\bibitem{gaurav2003}
G.~Sharma, \emph{Digital Color Imaging: Handbook}, ser. The Electrical Engineering and Applied Signal Processing Series.\hskip 1em plus 0.5em minus 0.4em\relax CRC Press, 2003.

\bibitem{ColorVideoVDP}
\BIBentryALTinterwordspacing
R.~K. Mantiuk, P.~Hanji, M.~Ashraf, Y.~Asano, and A.~Chapiro, ``Colorvideovdp: A visual difference predictor for image, video and display distortions,'' \emph{ACM Trans. Graph.}, vol.~43, no.~4, jul 2024. [Online]. Available: \url{https://doi.org/10.1145/3658144}
\BIBentrySTDinterwordspacing

\bibitem{David1988}
H.~A. David, \emph{The Method of Paired Comparisons, 2nd ed.}\hskip 1em plus 0.5em minus 0.4em\relax Oxford University Press, Educational Supply Section, North Kettering Business Park, Hipwell Road, Kettering, Northamptonshire. NN14 1UA, UK: Oxford University Press, 1988.

\bibitem{Banterle2009}
F.~Banterle, P.~Ledda, K.~Debattista, M.~Bloj, A.~Artusi, and A.~Chalmers, ``A psychophysical evaluation of inverse tone mapping techniques,'' \emph{Comput. Graph. Forum}, vol.~28, no.~1, pp. 13--25, 2009.

\bibitem{YeePattanaik}
Y.~H. Yee and S.~N. Pattanaik, ``Segmentation and adaptive assimilation for detail-preserving display of high-dynamic range images,'' \emph{The Visual Computer}, vol.~19, pp. 457--466, 2003.

\bibitem{AshikhminTMO}
M.~Ashikhmin, ``A tone mapping algorithm for high contrast images,'' in \emph{Proceedings of the 13th Eurographics Workshop on Rendering}, ser. EGRW '02.\hskip 1em plus 0.5em minus 0.4em\relax Goslar, DEU: Eurographics Association, 2002, p. 145–156.

\bibitem{BanterleTMO}
\BIBentryALTinterwordspacing
F.~Banterle, A.~Artusi, E.~Sikudova, T.~Bashford-Rogers, P.~Ledda, M.~Bloj, and A.~Chalmers, ``Dynamic range compression by differential zone mapping based on psychophysical experiments,'' in \emph{Proceedings of the ACM Symposium on Applied Perception}, ser. SAP '12.\hskip 1em plus 0.5em minus 0.4em\relax New York, NY, USA: Association for Computing Machinery, 2012, p. 39–46. [Online]. Available: \url{https://doi.org/10.1145/2338676.2338685}
\BIBentrySTDinterwordspacing

\bibitem{BruceTMO}
\BIBentryALTinterwordspacing
N.~D. Bruce, ``Expoblend: Information preserving exposure blending based on normalized log-domain entropy,'' \emph{Computers \& Graphics}, vol.~39, pp. 12--23, 2014. [Online]. Available: \url{https://www.sciencedirect.com/science/article/pii/S0097849313001428}
\BIBentrySTDinterwordspacing

\bibitem{ChiuTMO}
\BIBentryALTinterwordspacing
K.~Chiu, M.~Herf, P.~Shirley, S.~Swamy, C.~Wang, and K.~Zimmerman, ``Spatially nonuniform scaling functions for high contrast images,'' in \emph{Proceedings of Graphics Interface '93}, ser. GI '93.\hskip 1em plus 0.5em minus 0.4em\relax Toronto, Ontario, Canada: Canadian Human-Computer Communications Society, 1993, pp. 245--253. [Online]. Available: \url{http://graphicsinterface.org/wp-content/uploads/gi1993-34.pdf}
\BIBentrySTDinterwordspacing

\bibitem{DragoTMO}
\BIBentryALTinterwordspacing
F.~Drago, K.~Myszkowski, T.~Annen, and N.~Chiba, ``Adaptive logarithmic mapping for displaying high contrast scenes,'' \emph{Computer Graphics Forum}, vol.~22, no.~3, pp. 419--426, 2003. [Online]. Available: \url{https://onlinelibrary.wiley.com/doi/abs/10.1111/1467-8659.00689}
\BIBentrySTDinterwordspacing

\bibitem{DurandTMO}
\BIBentryALTinterwordspacing
F.~Durand and J.~Dorsey, ``Fast bilateral filtering for the display of high-dynamic-range images,'' \emph{ACM Trans. Graph.}, vol.~21, no.~3, p. 257–266, July 2002. [Online]. Available: \url{https://doi.org/10.1145/566654.566574}
\BIBentrySTDinterwordspacing

\bibitem{FattalTMO}
\BIBentryALTinterwordspacing
R.~Fattal, D.~Lischinski, and M.~Werman, ``Gradient domain high dynamic range compression,'' \emph{ACM Transactions on Graphics}, vol.~21, no.~3, p. 249–256, July 2002. [Online]. Available: \url{https://doi.org/10.1145/566654.566573}
\BIBentrySTDinterwordspacing

\bibitem{FerwerdaTMO}
\BIBentryALTinterwordspacing
J.~A. Ferwerda, S.~N. Pattanaik, P.~Shirley, and D.~P. Greenberg, ``A model of visual adaptation for realistic image synthesis,'' in \emph{Proceedings of the 23rd Annual Conference on Computer Graphics and Interactive Techniques}, ser. SIGGRAPH '96.\hskip 1em plus 0.5em minus 0.4em\relax New York, NY, USA: Association for Computing Machinery, 1996, p. 249–258. [Online]. Available: \url{https://doi.org/10.1145/237170.237262}
\BIBentrySTDinterwordspacing

\bibitem{KimKautzTMO}
M.~H. Kim and J.~Kautz, ``Consistent tone reproduction,'' in \emph{Proceedings of the Tenth IASTED International Conference on Computer Graphics and Imaging}, ser. CGIM '08.\hskip 1em plus 0.5em minus 0.4em\relax USA: ACTA Press, 2008, p. 152–159.

\bibitem{KrawczykTMO}
G.~Krawczyk, K.~Myszkowski, and H.-P. Seidel, ``Lightness perception in tone reproduction for high dynamic range images,'' \emph{Computer Graphics Forum}, 2005.

\bibitem{KuangTMO}
\BIBentryALTinterwordspacing
J.~Kuang, G.~M. Johnson, and M.~D. Fairchild, ``{iCAM}06: A refined image appearance model for {HDR} image rendering,'' \emph{Journal of Visual Communication and Image Representation}, vol.~18, no.~5, pp. 406--414, 2007, special issue on High Dynamic Range Imaging. [Online]. Available: \url{https://www.sciencedirect.com/science/article/pii/S1047320307000533}
\BIBentrySTDinterwordspacing

\bibitem{LiTMO}
\BIBentryALTinterwordspacing
Y.~Li, L.~Sharan, and E.~H. Adelson, ``Compressing and companding high dynamic range images with subband architectures,'' \emph{ACM Trans. Graph.}, vol.~24, no.~3, p. 836–844, July 2005. [Online]. Available: \url{https://doi.org/10.1145/1073204.1073271}
\BIBentrySTDinterwordspacing

\bibitem{LischinskiTMO}
\BIBentryALTinterwordspacing
D.~Lischinski, Z.~Farbman, M.~Uyttendaele, and R.~Szeliski, ``Interactive local adjustment of tonal values,'' \emph{ACM Trans. Graph.}, vol.~25, no.~3, p. 646–653, July 2006. [Online]. Available: \url{https://doi.org/10.1145/1141911.1141936}
\BIBentrySTDinterwordspacing

\bibitem{MertensTMO}
T.~Mertens, J.~Kautz, and F.~Van~Reeth, ``Exposure fusion,'' in \emph{15th Pacific Conference on Computer Graphics and Applications (PG'07)}, October 2007, pp. 382--390.

\bibitem{RamanTMO}
S.~Raman and S.~Chaudhuri, ``{Bilateral Filter Based Compositing for Variable Exposure Photography},'' in \emph{Eurographics 2009 - Short Papers}, P.~Alliez and M.~Magnor, Eds.\hskip 1em plus 0.5em minus 0.4em\relax The Eurographics Association, 2009.

\bibitem{ReinhardTMO}
\BIBentryALTinterwordspacing
E.~Reinhard, M.~Stark, P.~Shirley, and J.~Ferwerda, ``Photographic tone reproduction for digital images,'' \emph{ACM Trans. Graph.}, vol.~21, no.~3, p. 267–276, July 2002. [Online]. Available: \url{https://doi.org/10.1145/566654.566575}
\BIBentrySTDinterwordspacing

\bibitem{ReinhardDevlinTMO}
E.~Reinhard and K.~Devlin, ``Dynamic range reduction inspired by photoreceptor physiology,'' \emph{IEEE Transactions on Visualization and Computer Graphics}, vol.~11, no.~1, pp. 13--24, January 2005.

\bibitem{SchlickTMO}
C.~Schlick, ``Quantization techniques for visualization of high dynamic range pictures,'' in \emph{Photorealistic Rendering Techniques}, G.~Sakas, S.~M{\"u}ller, and P.~Shirley, Eds.\hskip 1em plus 0.5em minus 0.4em\relax Berlin, Heidelberg: Springer Berlin Heidelberg, 1995, pp. 7--20.

\bibitem{TumblinTMO}
\BIBentryALTinterwordspacing
J.~Tumblin, J.~K. Hodgins, and B.~K. Guenter, ``Two methods for display of high contrast images,'' \emph{ACM Trans. Graph.}, vol.~18, no.~1, p. 56–94, January 1999. [Online]. Available: \url{https://doi.org/10.1145/300776.300783}
\BIBentrySTDinterwordspacing

\bibitem{VanHaterenTMO}
J.~H. van Hateren, ``\BIBforeignlanguage{English}{Encoding of high dynamic range video with a model of human cones},'' \emph{\BIBforeignlanguage{English}{Acm transactions on graphics}}, vol.~25, no.~4, pp. 1380--1399, October 2006, relation: http://www.rug.nl/informatica/organisatie/overorganisatie/iwi Rights: University of Groningen. Research Institute for Mathematics and Computing Science (IWI).

\bibitem{WardTMO}
\BIBentryALTinterwordspacing
G.~Ward~Larson, H.~Rushmeier, and C.~Piatko, ``A visibility matching tone reproduction operator for high dynamic range scenes,'' in \emph{ACM SIGGRAPH 97 Visual Proceedings: The Art and Interdisciplinary Programs of SIGGRAPH '97}, ser. SIGGRAPH '97.\hskip 1em plus 0.5em minus 0.4em\relax New York, NY, USA: Association for Computing Machinery, 1997, p. 155. [Online]. Available: \url{https://doi.org/10.1145/259081.259242}
\BIBentrySTDinterwordspacing

\bibitem{WardGlobalTMO}
\BIBentryALTinterwordspacing
G.~Ward, ``A contrast-based scalefactor for luminance display,'' in \emph{Graphics Gems}, P.~S. Heckbert, Ed.\hskip 1em plus 0.5em minus 0.4em\relax Elsevier, 1994, pp. 415--421. [Online]. Available: \url{https://doi.org/10.1016/b978-0-12-336156-1.50054-9}
\BIBentrySTDinterwordspacing

\end{thebibliography}

\begin{IEEEbiography}
[{\includegraphics[width=1in,height=1.25in,clip,keepaspectratio]{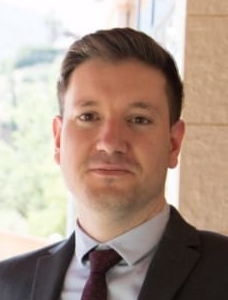}}]
{Xenios Milidonis}
received a B.Sc. in Physics from the University of Cyprus in 2011, an M.Sc. in Physics and Computing in Medicine and Biology (Medical Physics) from the University of Manchester in 2012, and a Ph.D. in Molecular and Clinical Medicine (Neuroimaging) from the University of Edinburgh in 2017. Since then, he worked as a post-doctoral Research Associate at the School of Biomedical Engineering \& Imaging Sciences, King’s College London, focusing on the fabrication of multimodality physical standards for the assessment and validation of dynamic cardiac first-pass perfusion imaging, as well as the development of algorithms for the automated quantitative analysis of magnetic resonance imaging data. He is currently a Senior Research Associate in the DeepCamera group at CYENS Centre of Excellence in Cyprus, working towards the development of deep learning-based tools for image processing, visualization and evaluation, while co-managing the DeepCamera group. He is also a registered Medical Physicist in Cyprus and a Visiting Scholar at King’s College London and the University of Cyprus.
\end{IEEEbiography}

\begin{IEEEbiography}
[{\includegraphics[width=1in,height=1.25in,clip,keepaspectratio]{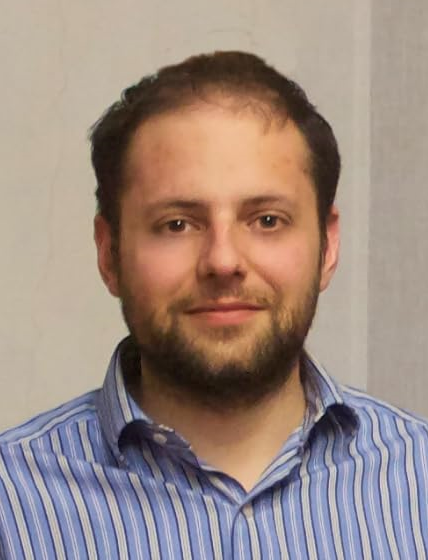}}]
{Francesco Banterle} is a full-time researcher at the Visual Computing Laboratory at ISTI-CNR, Italy. He received a Ph.D. in Engineering from Warwick University in 2009 when he developed Inverse Tone Mapping that bridges the gap between SDR and HDR Imaging.
He is the first author of the book ``Advanced High Dynamic Range Imaging'', a reference book for HDR imaging research, and co-author of the book ``Image Content Retargeting''. His main research interests are in the field of HDR imaging, Computer Graphics (image-based lighting), Computer Vision, and Deep Learning.
\end{IEEEbiography}

\begin{IEEEbiography}
[{\includegraphics[width=1in,height=1.25in,clip,keepaspectratio]{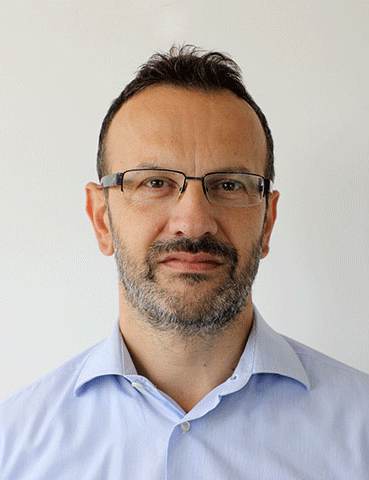}}]
{Alessandro Artusi} received the Ph.D. degree in computer science from the Vienna University of Technology, in 2004. He is currently a Research Associate Professor and the Managing Director of the DeepCamera group at CYENS Centre of Excellence in Cyprus. Recently, he joined as a Founding Member of Moving Picture, Audio and Data Coding by Artificial Intelligence (MPAI), a non-profit standards organization established in Geneva. His research interests include visual perception, image/video processing, HDR technology, objective/subjective imaging/video evaluation, deep learning, computer vision, and color science. He was a recipient of the prestigious BSI Award for his work on the JPEGXT standard. He was appointed as the Head of Delegation for the Cypriot National Body for the ISO/IEC/JCT 1 SC29 Committee. In the past, he has been a BSI Member of the IST/37 Committee and the UK Representative for the JPEG and MPEG Standardization Committees.
\end{IEEEbiography}

\end{document}